\documentclass[conference]{IEEEtran}
\IEEEoverridecommandlockouts
\usepackage{cite}
\usepackage{amsmath,amssymb,amsfonts}
\usepackage{tabularx}
\usepackage{graphicx}
\usepackage{textcomp}
\usepackage{balance}
\usepackage{todonotes}
\usepackage[table]{xcolor}
\usepackage{booktabs}
\usepackage{multirow}
\usepackage{url}
\makeatletter
\newcommand{\newlineauthors}{%
  \end{@IEEEauthorhalign}\hfill\mbox{}\par
  \mbox{}\hfill\begin{@IEEEauthorhalign}
}
\makeatother
\def\BibTeX{{\rm B\kern-.05em{\sc i\kern-.025em b}\kern-.08em
    T\kern-.1667em\lower.7ex\hbox{E}\kern-.125emX}}
\begin{document}

\title{Adoption-Ready Project-Based Learning for Computing Education: The FORAP Framework and a Multi-Scale Project Portfolio}

\author{
    \IEEEauthorblockN{Ahmad D. Suleiman}
    \IEEEauthorblockA{
        \textit{Computing and Information Sciences} \\
        \textit{Rochester Institute of Technology}\\
        Rochester, NY, USA \\
        as4300@rit.edu
    }
    
    \and
    
    \IEEEauthorblockN{Jan DeWaters}
    \IEEEauthorblockA{
        \textit{Institute for STEM Education}\\
        \textit{Clarkson University}\\
        Potsdam, NY, USA \\
        jdewater@clarkson.edu
    }
    
    \and
    
    \IEEEauthorblockN{David C Shepherd}
    \IEEEauthorblockA{
        \textit{Computer Science}\\
        \textit{Louisiana State University}\\
        Baton Rouge, LA, USA \\
        dshepherd@lsu.edu
    }
    
    \and
    
    \IEEEauthorblockN{Turgay Korkmaz}
    \IEEEauthorblockA{
        \textit{Computer Science}\\
        \textit{UT San Antonio}\\
        San Antonio, TX, USA \\
        Turgay.Korkmaz@utsa.edu
    }
    
    \and
    
    \IEEEauthorblockN{Faraz Hussain}
    \IEEEauthorblockA{
        \textit{Electrical \& Computer Engineering}\\
        \textit{Clarkson University}\\
        Potsdam, NY, USA \\
        fhussain@clarkson.edu
    }
    
    \and
    
    \IEEEauthorblockN{Yu Liu}
    \IEEEauthorblockA{
        \textit{Electrical \& Computer Engineering}\\
        \textit{Clarkson University}\\
        Potsdam, NY, USA \\
        yuliu@clarkson.edu
    }
    
    \and
    
    \IEEEauthorblockN{Daqing Hou}
    \IEEEauthorblockA{
        \textit{Software Engineering} \\
        \textit{Rochester Institute of Technology}\\
        Rochester, NY, USA \\
        dqvse@rit.edu
    }
}

\IEEEaftertitletext{\vspace{-2em}}

\maketitle

% Project-based learning (PjBL)  
\begin{abstract}
This innovative practice full paper presents FORAP (Framework for Organizing Reusable and Adaptable PjBL Projects) and a portfolio of 14 adoption-ready project-based learning (PjBL) project packages built with the framework. PjBL in computing education offers strong educational benefits, yet its adoption remains limited by high instructor workload and recurring student technical challenges. FORAP addresses these barriers by organizing each package around a project designed with aligned learning objectives and described through project attributes, along with coordinated instructor, student, and assessment materials that support adoption and adaptation across diverse computing courses.
We report on four years of deployment across 44 classroom trials at seven universities, drawing on feedback from students, instructors, and advisory board members. Results suggest that structured project packaging supports feasible adoption with limited modification effort and that targeted support materials help reduce the technical barriers that commonly hinder student engagement.
The contributions of this work include FORAP and a multi-scale portfolio that demonstrates its use across diverse computing domains and project scopes, offering practical guidance for instructors who wish to design, adopt, or adapt reusable PjBL projects in computing education.

\end{abstract}

\begin{IEEEkeywords}
Project based learning, Course design, adoption
\end{IEEEkeywords}

\vspace{-1em}
\section{Introduction}
\label{section:introduction}

Project-based Learning (PjBL) is widely used in computing education to provide authentic, hands-on learning experiences, but implementing and sustaining it in practice remains challenging for both instructors and students~\cite{suleiman2025factors,suleiman2024providing}.
For instructors, substantial barriers arise when adopting and maintaining PjBL projects, including the time and effort required to design new projects or adapt existing ones, align them with course objectives, scope them appropriately, manage project organization in class, support student teams, and assess open-ended project work fairly~\cite{suleiman2025factors}. These barriers are often exacerbated by limited institutional support, such as insufficient time, training, teaching assistants, funding, and reusable project resources. For students,  recurring technical barriers arise during project work, including installation and configuration of tools, lack of prerequisite knowledge, and difficulties while completing project tasks such as debugging, implementation, testing, deployment, version control, and documentation~\cite{suleiman2024providing}. When students perceive these challenges as insurmountable, their motivation and confidence can weaken, especially when they feel stuck and cannot move forward.

This paper addresses these barriers through two contributions: FORAP (Framework for Organizing Reusable and Adaptable PjBL Projects) and a multi-scale portfolio of 14 adoption-ready computing PjBL project packages. FORAP organizes each project as a coordinated package centered on aligned learning objectives, described through project attributes, and supported by instructor, student, and assessment materials. Multi-scale refers to projects that vary in scope, complexity, domain, and codebase context.
The portfolio demonstrates how this packaging can support reuse and adaptation across diverse computing areas and technologies, including software engineering, cybersecurity, machine learning, behavioral biometrics, and databases.

Our work builds on prior work in PjBL pedagogy and reusable computing education resources. It is grounded in established PjBL principles~\cite{thomas2000} and uses aligned learning objectives as the organizing anchor for project packaging~\cite{suleiman2023mapping}. Prior frameworks such as Gold Standard PBL and xPBL provide useful guidance for designing and managing PjBL experiences~\cite{larmer2015gold,dos2014xpbl}, while resources such as Nifty Assignments and ROSE support reuse by sharing assignments, projects, and education-friendly open-source systems~\cite{niftyAssignments,meneely2008rose}. However, these frameworks and resources do not typically organize projects as coordinated, adoption-ready packages that combine instructor, student, and assessment support structures. FORAP directly addresses these gaps by making reuse and adaptation explicit through coordinated packaging.

We evaluate this innovative practice through evidence gathered over four years of design, refinement, and classroom deployment of FORAP project packages. This evidence includes iterative feedback from advisory board review, student beta testing, student and instructor post-implementation feedback, student pre- and post-surveys, and deployment outcomes across classroom settings. 
Across classroom trials, FORAP-packaged projects proved feasible to adopt with limited adaptation effort, while the included support materials helped reduce common technical barriers.

This paper focuses on presenting the multi-scale project portfolio, the framework used to create and package these projects, the evaluation process, and lessons learned from their classroom trials.
It makes two contributions:
\begin{enumerate}
    \item \textbf{FORAP:} a framework for organizing and packaging computing PjBL projects for reuse and adaptation.
    \item \textbf{Portfolio:} a set of diverse adoption-ready projects that demonstrate how to apply the framework.
\end{enumerate}

The remainder of the paper is organized as follows. Section~\ref{section:background_motivation} presents the instructor and student imperatives that motivate this work. Section~\ref{section:prior_work} connects our work to PjBL principles, PjBL design frameworks, and reusable computing project repositories. Section~\ref{section:forap} presents FORAP, and Section~\ref{section:portfolio} describes the multi-scale project portfolio. Section~\ref{section:evaluation_methods} describes the evaluation methods. Section~\ref{section:outcomes_lessons} reports outcomes and lessons learned. Section~\ref{section:discussion_implications} discusses implications and challenges ahead. Section~\ref{section:conclusion} concludes the paper.

\section{Background and Motivation}
\label{section:background_motivation}
\subsection{Instructor Imperative}
A central motivation for this work is the effort required from instructors to adopt, implement, and sustain PjBL projects over time. Prior work shows that instructors face substantial workload in designing new projects, adapting existing ones, aligning projects with course objectives, organizing and managing the learning process, supporting students during project work, and assessing open-ended outcomes~\cite{suleiman2025factors}. Project scoping is a persistent challenge, and even when instructors choose to reuse an existing project rather than create a new one, adoption still requires time to review the project, scope it to the course context, plan student support, and prepare an appropriate assessment. These barriers make sustained PjBL implementation difficult, even for instructors who value the approach and want to continue using it~\cite{suleiman2025factors}.

\subsection{Student Imperative}
A second motivation is the recurring technical challenges students face during project work. A prior study has identified three common categories: installation and configuration of required tools, lack of prerequisite knowledge, and challenges while completing project tasks such as debugging, implementation, testing, deployment, version control, and documentation~\cite{suleiman2024providing}. These challenges can slow progress, frustrate students, and weaken motivation when perceived as insurmountable. They also increase the amount of support instructors must provide, especially in large classes and in settings with diverse student backgrounds and technologies~\cite{suleiman2024providing}. This creates an imperative to provide project structure and support that helps students keep moving forward during project work.

\section{Connection to Prior Work}
\label{section:prior_work}

Prior work in computing education has explored multiple aspects of PjBL, including pedagogical principles, design frameworks for managing project-based courses, and repositories that support reuse of computing assignments and projects. This section situates our work within these areas.

\subsection{PjBL Principles}
Prior work characterizes PjBL as an instructional approach in which students learn through meaningful projects carried out over extended periods of time~\cite{thomas2000}. Building on this foundation, we focus on three principles that are especially relevant to our work:
\begin{itemize}
    \item \textbf{Authentic learning:} projects connect course concepts to realistic contexts, tasks, and practices.
    \item \textbf{Collaborative learning:} projects create opportunities for teamwork, communication, and shared responsibility.
    \item \textbf{Self-directed learning:} projects create opportunities for students to learn, plan, research, make decisions, and work through ambiguity with appropriate support.
\end{itemize}

\subsection{PjBL Design Frameworks}
Gold Standard PBL~\cite{larmer2015gold} and xPBL~\cite{dos2014xpbl} provide guidance for designing and managing PjBL experiences. These frameworks are useful for thinking about project design and classroom implementation, including what kinds of support should be built into a project to help instructors and students during project work. However, they do not focus on organizing reusable and adaptable projects as adoption-ready packages, and they do not directly address the instructor and student challenges summarized in Section~\ref{section:background_motivation}.

\subsection{Reusable Computing Projects and Repositories}
Prior work has also supported reuse by sharing computing assignments and projects. Nifty Assignments curates classroom-tested assignments, typically smaller in scale and intended for direct classroom adoption across computing topics~\cite{niftyAssignments}. ROSE provides education-friendly open-source projects intended for instructional use and reuse~\cite{meneely2008rose}. Open-source projects can also support brownfield experiences, where students engage with large, non-trivial existing codebases that better resemble real software systems~\cite{pinto2017training}. These resources provide valuable starting points for reuse, but they do not typically organize projects as coordinated adoption-ready packages that include structured instructor, student, and assessment support.

\subsection{Gaps and Challenges Addressed by FORAP}

Motivated by the instructor and student imperatives in Section~\ref{section:background_motivation}, we focus on three gaps that limit the reuse and adaptation of PjBL projects across course contexts. First, prior work highlights the need for student-facing support materials that address installation and configuration, prerequisite gaps, and common project-task difficulties~\cite{suleiman2024providing}. Second, instructors need stronger support materials that reduce the workload of project adoption, implementation, and sustainment over time~\cite{suleiman2025factors}. Third, reusable projects often lack formative assessment structures, such as clear milestones, checkpoints, and aligned assessment artifacts, that can support student progress while also reducing instructor workload~\cite{suleiman2025factors}.

Taken together, these gaps point to the need for PjBL projects that are easier for instructors to adopt and adapt while also providing structured support to students throughout project work. These observations motivated the following design goals for FORAP:

\subsubsection{Aligned Learning Objectives} Center project design on clear learning objectives that describe what students are expected to know and be able to do after completing the project. These objectives guide the selection and organization of project tasks and support materials. Clearly defined objectives support constructive alignment among learning goals, project activities, and assessment~\cite{kandlbinder2014constructive}.

\subsubsection{Reusability and Adaptability} Package projects in a modular and flexible form that supports reuse across offerings and adaptation to different course contexts and goals.

\subsubsection{Support for Teaching and Learning} Include support materials that address instructor barriers of PjBL adoption~\cite{suleiman2025factors} and address common student technical challenges~\cite{suleiman2024providing}.

\subsubsection{Formative and Summative Assessment} Include structured milestones, checkpoints, and aligned assessment artifacts to support formative feedback during project work and summative evaluation of student learning.

These design goals informed the FORAP components described in Section~\ref{section:forap}.

\section{FORAP and Its Components}
\label{section:forap}

\begin{figure}[ht]
    \centering
    \includegraphics[width=\linewidth]{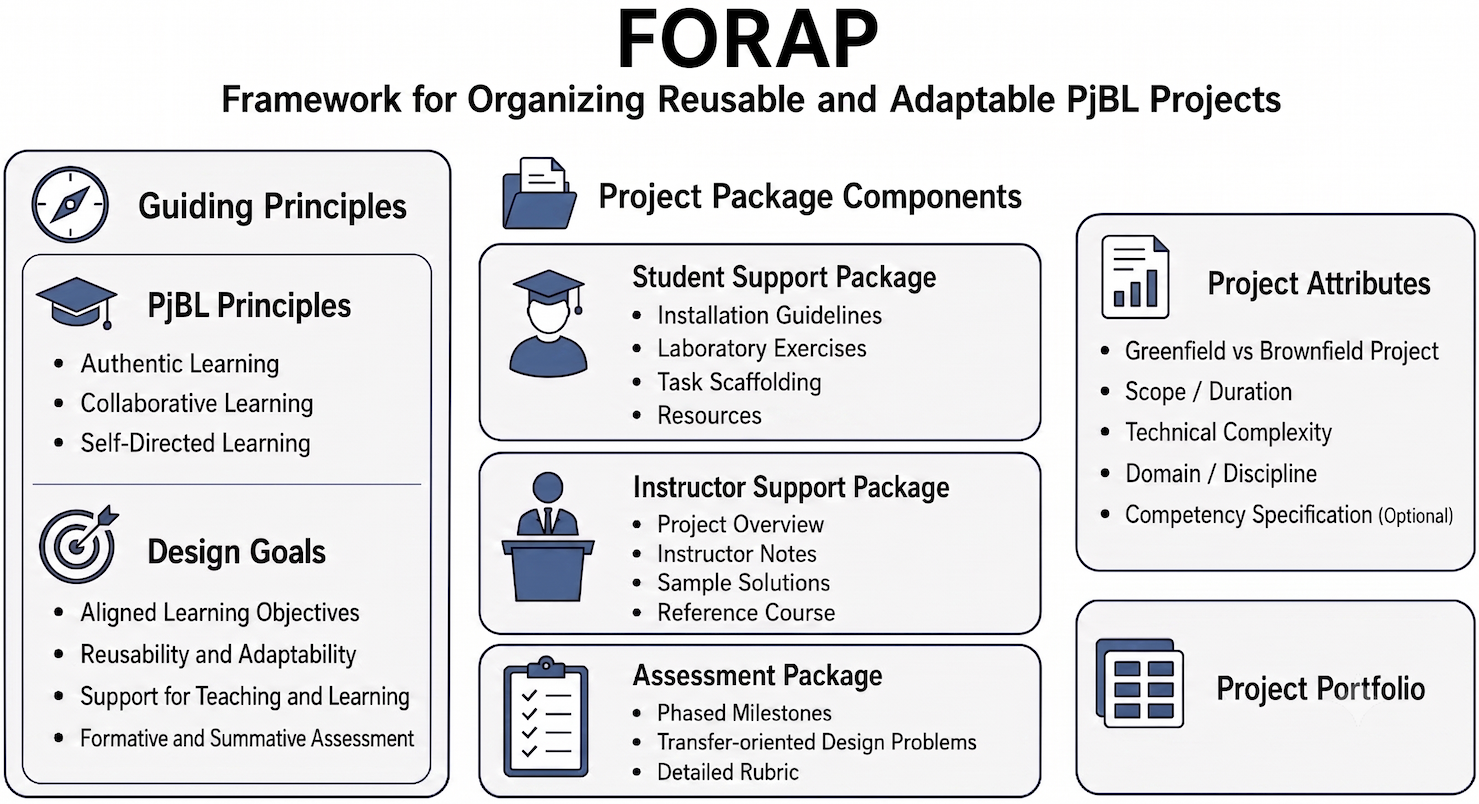}
    \caption{FORAP Architecture}
    \label{fig:forap}
\end{figure}

Figure~\ref{fig:forap} shows FORAP’s architecture and its components. The framework is guided by PjBL principles and the design goals. Each project includes three coordinated support packages: instructor, student, and assessment. In addition, each package is described by a set of project attributes, which makes projects easier to compare and adapt. 

 % \vspace{-1em}

\subsection{Project Attributes}
Each FORAP project is described through a set of attributes that help instructors understand its nature, fit, and adaptability. These attributes provide a concise project profile that supports comparison across projects, helps instructors judge course fit, and identifies where adaptation may be needed for different contexts, constraints, and instructional goals.

\subsubsection{\textbf{Greenfield vs. Brownfield Project}} whether the project starts from a new codebase that students build from scratch or from an existing system that students modify, extend, test, refactor, or maintain.

\subsubsection{\textbf{Scope and Duration}} the expected size of the project and the time required for completion.

\subsubsection{\textbf{Technical Complexity}} the expected level of difficulty based on implementation effort, number of technical components and dependencies, setup burden, integration across technologies, and whether students work in an existing codebase.

\subsubsection{\textbf{Domain/Discipline}} the computing areas or subject contexts in which the project is situated.

\subsubsection{\textbf{Competency Specification}} FORAP optionally supports mapping of learning objectives~\cite{suleiman2023mapping} to a curriculum standard, such as the competency specification of CC2020~\cite{CC2020}. This helps instructors compare objectives across projects, courses, and programs. Bloom's Taxonomy can be used to characterize the cognitive level of each objective, helping clarify the thinking and performance a project is intended to develop~\cite{bloom}.

\subsection{The Instructor Support Package}
The Instructor Support Package includes the materials needed to help instructors adopt, implement, and adapt a project with less effort. It is intended to reduce common barriers related to time, planning, project organization, student support, and long-term sustainment~\cite{suleiman2025factors}. To support this goal, the package may include the following:
\begin{itemize}
    \item \textbf{Project Overview}: A summary of the project’s learning objectives, attributes, nature of tasks, and guidance for adopting and adapting the project, including scope, alignment and considerations for local course contexts.
    \item \textbf{Instructor Notes}: Notes that support project implementation and adaptation, including past deployment experiences, sample schedules, alternative project designs, common student challenges, and guidance for instructional decisions during project use. These notes also identify files, dependencies, tools, setup instructions, and services that may need updates as technologies change.
    \item \textbf{Sample Solutions}: Example solutions for project tasks similar to those expected from students, which can support instructor preparation, demonstration, and evaluation.
    \item \textbf{Reference Course}: Examples of past project deployments in courses that help instructors understand how the project has been integrated into the syllabus, course schedule, assessment, and classroom use in practice.
\end{itemize}

\subsection{The Student Support Package}
The Student Support Package includes the materials provided to help students make progress during project work. It is designed to address common technical challenges students face, especially installation/configuration, prerequisite gaps, and difficulties encountered while completing project tasks~\cite{suleiman2024providing}. To support this goal, the package may include the following:
\begin{itemize}
    \item \textbf{Installation Guidelines}: Materials that help students install, configure, and access the required tools and environments for the project. Where applicable, these may also include installation scripts and pre-configured environments such as virtual machines or Docker containers to reduce setup effort and improve consistency.
    
    \item \textbf{Laboratory Exercises}: Targeted exercises that help students build prerequisite knowledge and prepare for project tasks, such as learning to use a programming or markup language, GitHub, a database system, or a language framework.
    
    \item \textbf{Task Scaffolding}: Project-specific scaffolding that students use directly to complete tasks, such as task instructions, milestone checklists, starter code, task-specific troubleshooting guidance, diagrams that guide implementation decisions, solution hints, and templates. The level of scaffolding may vary from step-by-step directions to basic task requirements.
    
    \item \textbf{Resources}: Reference materials students consult as needed for background understanding, such as official documentation for APIs, languages, and frameworks, report templates, example artifacts, and other resources.
\end{itemize}

\subsection{The Assessment Package}
The Assessment Package includes materials used to support formative and summative assessment in a FORAP project. It is organized around the project's learning objectives so that assessment remains connected to what students are expected to know and be able to do \cite{kandlbinder2014constructive}. The package includes:

\begin{itemize}
    \item \textbf{Phased milestones}: Structured milestones and checkpoints that define intermediate deliverables and enable formative assessment during project execution, including feedback points tied to milestone artifacts.
    \item \textbf{Transfer-oriented design problems}\cite{suleiman2026llm}: A set of scenario-based problems that assess learning beyond the original project and can be deployed in exams or presentations to elicit higher-order thinking (analysis, evaluation, synthesis) aligned with Bloom's Taxonomy \cite{bloom}.
    \item \textbf{Detailed rubric}: A grading rubric aligned with the learning objectives, with clear criteria for evaluating project artifacts such as code, design, reports, and presentations.
\end{itemize}

\section{Multi-Scale Project Portfolio}
\label{section:portfolio}
This paper contributes a multi-scale portfolio of 14 computing PjBL projects packaged using FORAP. The projects were developed over multiple years from multiple sources consistent with~\cite{suleiman2025factors}, including course needs identified by instructors, faculty domain expertise, by-products of completed research projects, open-source projects, and prior classroom student project ideas. They were then refined through the quality assurance process described in Section~\ref{section:evaluation_methods}. The portfolio includes projects that vary in scope/duration, technical complexity, domain/discipline, and mix of greenfield and brownfield projects. Table~\ref{tab:portfolio-projects} summarizes these project attributes and includes classroom trial counts for each project. All 14 projects were organized using the FORAP package structure described in Section~\ref{section:forap}, including instructor, student, and assessment support, with project-specific artifacts tailored to each project.

\begin{table*}[htb]
\centering
\small
\caption{High-level overview of the FORAP project portfolio.}
\label{tab:portfolio-projects}
\renewcommand{\arraystretch}{1.25}
\rowcolors{2}{gray!10}{white}
\begin{tabularx}{\textwidth}{p{0.12\textwidth} | p{0.12\textwidth} p{0.07\textwidth} p{0.075\textwidth} p{0.056\textwidth} X | p{0.01\textwidth}}
\toprule
\textbf{Project} & \textbf{Domain/\newline Discipline} & \textbf{Codebase Context} & \textbf{Technical \newline Complexity} & \textbf{Scope/ \newline Duration} & \textbf{Representative Learning Objective/ \newline Competency Statement} & \textbf{$N$} \\
\midrule
Breached! & Security, Database, Web & Greenfield & Low & 1--2 weeks & Design a solution for securing personal data by applying concepts of hashing. & 8 \\
Web Authentication System & Security, Web & Greenfield & Low & 4 weeks & Implement core user authentication features e.g., registration, login, secure password \& session management. & 3 \\
Authentication \& Authorization (OAuth) & Security, Web & Greenfield & Medium & 4--6 weeks & Implement an OAuth 2.0/OIDC workflow by building a client, integrating with an external provider, and creating a custom provider. & 3 \\
Touchalytics & Biometrics, ML, Mobile & Greenfield & High & 6--8 weeks & Apply machine learning to behavioral data classification for mobile authentication. & 5  \\
Distributed Tic-Tac-Toe & Distributed Systems, Mobile & Greenfield & Medium & 6--8 weeks & Design a client-server Tic-Tac-Toe game for mobile devices using socket programming and multithreading. & 2 \\
GPU & Parallel Computing & Greenfield & Medium & 4 weeks & Implement and evaluate GPU-based parallel processing for large-scale image manipulation. & 1 \\
Mango & Maintenance, Web & Brownfield & Medium & 6--8 weeks & Apply the software change process to modify and improve an open-source system. & 7  \\
Moodle & Maintenance, Web, LMS & Brownfield & High & 8--12 weeks & Apply refactoring, testing, and impact analysis to modify and extend an open-source system. & 3 \\
RSA & Cryptography & Greenfield & Low & 1--2 weeks & Implement and test RSA encryption and decryption, then debug a faulty RSA implementation.& 6  \\
Soteria & Web, Biometrics & Brownfield & Medium & 4--6 weeks & Identify and modify existing code based on feature descriptions and new requirements. & 1 \\
University Management & Databases, Web & Brownfield & Medium & 4--6 weeks & Develop a web system that uses and extends an existing database on course enrollment and faculty. & 1 \\
VS Code & Maintenance, Extension & Brownfield & High & 6--8 weeks & Modify and extend an existing codebase by implementing specified enhancements. & 1 \\
Web Server & Systems & Greenfield & Medium & 4--6 weeks & Design and implement a web server that supports core request-handling functionality. & 1 \\
Xfig & Legacy System, Graphics & Brownfield & High & 6--10 weeks & Modify and extend an existing software system by implementing specified enhancements and fixes. & 2 \\
\bottomrule
\end{tabularx}

\vspace{2pt}
\noindent\parbox{\textwidth}{\footnotesize\textit{Note:} The FORAP project package portfolio is available at \url{https://forap-docs.github.io/portfolio.html}. Upon request, access to the relevant project folders will be made available to qualified instructors. $N$ is the number of classroom trials.}

\end{table*}

\subsection{Portfolio Design Dimensions}
The portfolio was intentionally designed to vary along several FORAP project attributes that affect how projects can be adopted for use in different course settings:
\begin{itemize}
    \item \textbf{Scope and Duration}: projects range from shorter modules of about four weeks to larger multi-task projects that can span a full semester.
    
    \item \textbf{Technical Complexity}: projects vary in technical depth and integration effort. From self-contained tasks with fewer components and dependencies, such as \textit{Web Authentication System} and \textit{RSA}, to larger projects that require integrating multiple technologies, working across broader systems, or modifying existing open-source codebases such as \textit{Touchalytics} and \textit{Moodle}.

    \item \textbf{Codebase Context}: the portfolio includes both greenfield projects and brownfield projects that require modifying an existing codebase or system.
    
    \item \textbf{Domain/Discipline}: projects are designed for deployment across undergraduate computing courses, including software engineering, cybersecurity, systems, machine learning, databases, and application areas such as behavioral biometrics, web, and mobile development.
\end{itemize}

\subsection{Illustrative Example: Touchalytics Project Package}

To illustrate how FORAP structures a project package, we use the Touchalytics project as an example. Touchalytics asks students to build an Android authentication application that uses swipe behavior as a biometric signal, stores interaction data in cloud storage, and processes it with a Python-based machine learning backend. The package also illustrates FORAP's optional competency specification by mapping its learning objective to CC2020 knowledge areas such as intelligent systems, data and information management, security technology and implementation, and collaboration and teamwork, with Bloom/skill levels ranging from applying to creating and dispositions such as collaborative, self-directed, and inventive.

\subsubsection{Task structure} The project is organized into four phased tasks that build on one another. Students build a swipe-enabled Android app, extract and save swipe data to Firebase, implement a Python backend for behavioral biometrics, and integrate swipe data with the Python backend for prediction. These phased tasks provide intermediate deliverables and support formative assessment before students complete the final integrated system.

\subsubsection{Instructor Support} Includes a project overview, sample solutions, instructor notes, and a reference course schedule. These materials help instructors plan deployment, guide teams through the mobile-cloud-ML workflow, support team check-ins, address common student challenges, and maintain the project when tools or platforms change.

\subsubsection{Student Support} Includes installation guidelines for Python, Android Studio, and Firebase, laboratory exercises for Android, Python setup, and Firebase configuration, task scaffolding for Android app development, Firebase database integration, Python backend development, and machine learning integration, and resources on behavioral biometrics, machine learning, Android, Firebase, and project reporting.

\subsubsection{Assessment Support} Includes phased milestones, transfer-oriented design problems, and a detailed rubric. Together, these support formative assessment through setup labs, staged milestones, weekly check-ins, and intermediate deliverables. They also support summative and transfer assessment through the final app behavior, backend prediction, presentation, rubric, and transfer-oriented design problems.

The Touchalytics package shows how the four project tasks align with the learning objectives by moving from swipe capture to data storage, backend processing, and prediction. It also illustrates how setup guides, labs, references, milestones, and rubrics support both the task sequence and the learning objectives. Its modular structure allows instructors to reuse or adapt the Android, Firebase, Python backend, machine learning, and integration portions independently, or to black-box selected parts, such as providing the machine learning code in a course focused on Android or a working Android app in a course focused on machine learning.

\section{Evaluation Methods}
\label{section:evaluation_methods}

\begin{figure*}[!tb]
    \centering
    \includegraphics[width=0.8\linewidth]{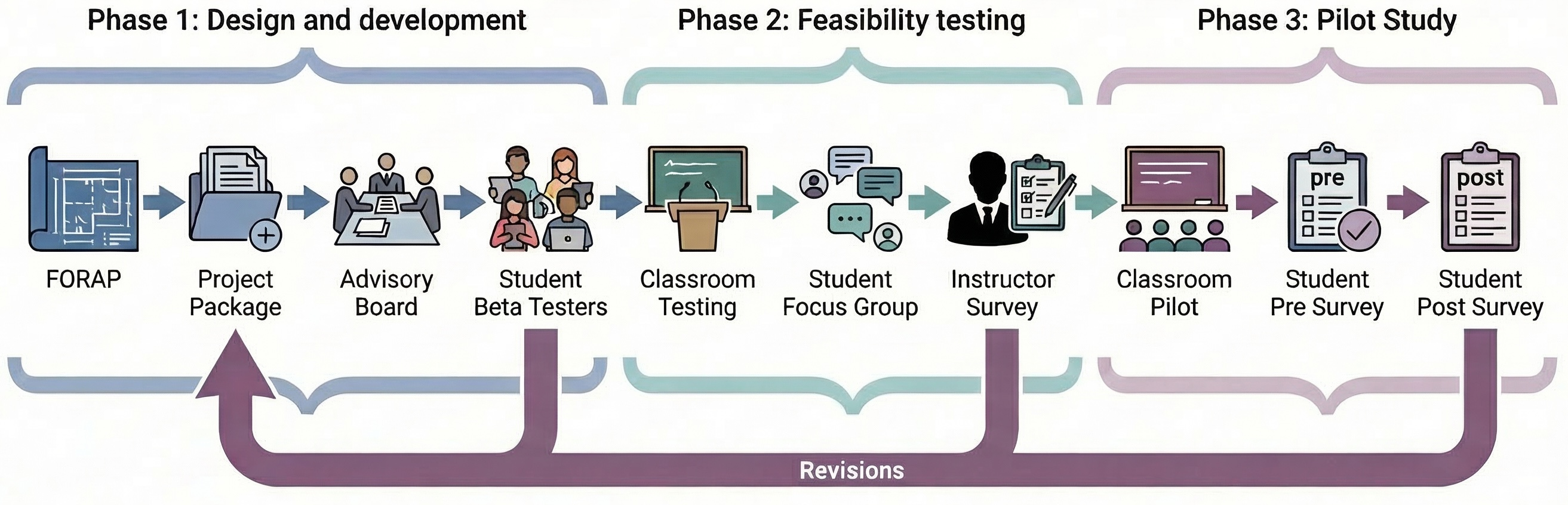}
    \caption{Quality assurance process for the design, development, refinement, and evaluation of the FORAP-based project portfolio.}
    \label{fig:portfolio-workflow}
\end{figure*}
% \vspace{-1em}

The portfolio projects were evaluated using evidence gathered during their design, development, refinement, and classroom trials over three project phases, as shown in Figure~\ref{fig:portfolio-workflow}. The three phases, consistent with the NSF/IES Common Guidelines for Design and Development research~\cite{common_guidelines_2013}, included:

\begin{itemize}
    \item \textbf{Phase 1: Design and Development.} Projects were packaged using FORAP, reviewed by an advisory board, and refined through student beta testing, with evidence from this phase used to revise the project package.
    \item \textbf{Phase 2: Feasibility Testing.} Projects underwent \textit{Classroom Testing} to examine implementation feasibility, and evidence from student focus groups and instructor feedback was used to guide further revisions.
    \item \textbf{Phase 3: Pilot Study.} More stable project packages underwent \textit{Classroom Pilot} with limited support to the instructor, and student pre- and post-surveys were included as part of the broader evaluation of student outcomes.
\end{itemize}
Our evaluation combined iterative feedback collected across these development phases with classroom project implementation. Feedback from beta testers, students, instructors, and advisory board members, described in the following sections, was reviewed to identify recurring themes and revision needs. 

\subsection{Advisory Board Review}
The advisory board served as an external expert review for both overall project direction and project-level revision. We typically engaged 2--4 domain experts to serve on an advisory board for each project, such as cybersecurity experts for Breached! project. Members reviewed project materials in advance, provided written comments, and then participated in a follow-up discussion to examine project rigor and relevance~\cite{daggett2005achieving}, authenticity, alignment with learning objectives, and the quality of support materials. We used this feedback to guide revisions to project packaging, task-to-objective alignment, instructor guidance, target course level, technology choices, and real-world authenticity.

\subsection{Student Beta Testers}
We typically engaged 2--4 undergraduate student beta testers for each project before it was tested in the classroom. Beta testers completed the project and then provided formative feedback through a post-survey, focus group, or interview. Their feedback helped us identify usability and scaffolding issues from a student perspective, including unclear goals, weak task progression, installation and configuration challenges, missing prerequisite support, unreasonable workload, and places where project tasks were difficult to complete. We used this feedback as an early feasibility check and to revise project structure, milestones, checkpoints, and support materials before classroom implementation.

\subsection{Student Focus Groups}
We used post-project student focus groups to gather qualitative feedback on student experience with the project after classroom implementation. These helped us understand aspects of the student experience that surveys could not fully capture, including clarity of goals, coherence across tasks, usefulness of support materials, communication channels, workload, teamwork, engagement, and perceived workforce relevance. The focus groups were conducted after project completion to ensure that participation would not impact project grades.

\subsection{Instructor Questionnaires and Focus Groups}
We developed instructor post-implementation instruments to collect both brief structured feedback and deeper qualitative feedback on project use. The instructor post-questionnaire is a brief measure that captures structured judgments and open-ended comments about project fit, rigor, student engagement, support materials, task coherence, authenticity, and likelihood of reuse. The instructor post-focus group, a deeper qualitative companion to the questionnaire, allowed us to explore how instructors (and their TAs) implemented and adapted the project, what challenges students and instructors faced, how much and what type of support were needed, and what revisions would improve future reuse. We used feedback from both instruments to guide iterative revisions to project packages.

\subsection{Student Pre and Post Questionnaires}
We developed a student pre- and post-survey instrument to examine changes in student outcomes, including personal competencies, during project participation. We used motivation as a key indicator of student outcomes because of its close relationship to engagement, self-confidence, self-efficacy, and academic success~\cite{motivation_edu,chemers2001academic,perkins_correlating}. The instrument was adapted from the Motivation Strategies for Learning Questionnaire (MSLQ)~\cite{pintrich1991manual}. To make repeated administration feasible, we used a reduced set of MSLQ items targeting motivational constructs relevant to PjBL, along with self-efficacy items adapted from~\cite{powers2015clics} and a small set of project-specific items about students' perceptions of the project. All items used a 5-point Likert-type scale. Although this instrument was part of the broader evaluation design, findings from this measure are not reported in this paper, whose main intent is to share the framework and portfolio of projects.

Detailed versions of all evaluation instruments used in this research will be made available upon request. 

\section{Outcomes and Lessons Learned}
\label{section:outcomes_lessons}

This section reports deployment outcomes and synthesizes recurring feedback from instructors and students across project use. We first summarize deployment counts over four project years and then present key lessons learned that informed iterative improvements to the project packages. 

\subsection{Classroom Implementation}
The 14 projects in our portfolio (Table~\ref{tab:portfolio-projects}) were each deployed at least once and overall used in 44 classroom trials across seven universities, including five in the United States and two in Canada, over four academic years. In all, 8 instructors, 22 courses, and about 2,000 students engaged with the projects. Some projects were used more than once across courses or years, so the number of classroom trials is larger than the number of distinct projects, instructors, and courses.

\subsection{Instructor Feedback}
\subsubsection{What worked well}
Across deployments, instructors were broadly positive about the projects. They consistently reported that the projects aligned well with course learning objectives, addressed realistic problems, and were worth using again in future offerings. Instructors also valued strong support packages, especially documentation, videos, laboratory exercises, and timely help from knowledgeable graduate assistants or TAs. These supports helped students bridge content gaps and helped faculty adapt the projects to their own courses. This feedback is especially important given the workload barriers reported by instructors, who noted that designing, adapting, supporting, and assessing PjBL projects requires substantial effort~\cite{suleiman2025factors}. By leveraging an existing FORAP package, instructors in our deployments were able to start with project documentation, student support materials, instructor guidance, and assessment resources already in place, rather than developing these components from scratch. Although we did not directly measure preparation time, anecdotal instructor feedback suggests that this packaging reduced the initial effort required to design, organize, adopt, and adapt PjBL activities.

\subsubsection{What needed improvement}
Instructor feedback across trials was used to inform subsequent revisions and later project deployment. Several projects needed better calibration: some required more scaffolding and clearer guidance, while others needed more open-ended problem-solving and critical thinking. Instructors often addressed this by revising materials, adding examples, introducing buggy code, or posing additional what-if questions. They also emphasized the need for greater modularity so that projects could be adjusted for course goals, student preparation, and available class time. In addition, instructors asked for stronger instructor-facing materials, including clearer notes, rubrics, milestone guidance, troubleshooting resources, and more explicit explanations of how project tasks map to course objectives.

\subsubsection{Persistent challenges}
Even when the project design itself was strong and improvements to the overall package had been made, some challenges persisted. Effective implementation still depended heavily on instructor preparation, TA readiness, class size, and platform or environment constraints. In some cases, grading remained complex, some tasks did not fit equally well across courses, and scaling project-based instruction to larger classes continued to require substantial staffing and coordination.

\subsection{Student Feedback}
\subsubsection{What worked well}
Students responded most positively when projects felt real, practical, and connected to authentic cybersecurity or software tasks. They especially valued projects that helped them see the big picture, apply prior knowledge, and work on tasks that resembled industry practice. Clear structure also mattered. Students responded more positively when projects were organized in a clear sequence, with manageable milestones and tasks that clearly built toward the final project. Support materials, office hours, and responsive instructional help also reduced barriers, especially when students were learning unfamiliar tools or technologies.

\subsubsection{What needed improvement}
Student feedback across trials was used to inform subsequent revisions and later project deployment. Student feedback repeatedly pointed to the need for stronger scaffolding in places where content was unfamiliar. Students asked for clearer setup guidance, more examples, simpler starter explanations, demonstrations for unfamiliar tools, and better troubleshooting support. Several concerns also centered on clarity, including unclear goals for some tasks, vague code explanations, confusing ordering of concepts, and materials that assumed too much prior knowledge. In a few projects, students also noted that the sequencing of lectures, labs, and tasks could be improved so that support appeared closer to the point of need.

\subsubsection{Persistent challenges}
Although many issues were addressed along the way, some issues were only partly resolved across deployments. Engagement persistently varied when projects were misaligned with students' interests, prior preparation, or course expectations. Teamwork was beneficial when well supported, but uneven participation and unclear expectations still reduced its effectiveness in some settings. Time and environment issues also remained a recurring challenge, as setup problems, outdated systems, and platform incompatibilities could consume substantial time.

\subsection{Lessons Learned and Resulting Revisions}
Evidence from classroom trials, instructor and student feedback, and advisory board review informed several recurring revisions to the project packages.

\subsubsection{Course placement and prerequisite readiness matter}
Projects were most successful when they matched students' prior knowledge and the intended point in the curriculum. As a result, project packages were revised to state prerequisite skills more clearly, identify appropriate course placements, and provide preparatory labs.

\subsubsection{Alignment must be explicit for both instructors and students}
Feedback showed that alignment was strongest when instructors could see how project tasks supported course learning objectives and when students could see why each task mattered. This led to clearer task-to-objective mapping, better distinctions between core and optional tasks, and stronger rubrics that link tasks, learning objectives, and assessment.

\subsubsection{Authenticity should be preserved and updated}
Real-world authenticity consistently supported engagement, but it was most effective when the technologies and scenarios felt current and recognizable. This led to revisions that retained realistic workplace and cybersecurity contexts while updating examples, tools, and project details where needed.

\subsubsection{Scaffolding must be balanced and coherent}
Too little scaffolding led to confusion, frustration, and overload, while too much reduced opportunities for critical thinking and problem solving. Therefore, revision focused not only on support level, but also on clearer problem statements, more coherent sequencing of tasks and labs, better milestones, and stronger explanations of how tasks build toward the overall project.

\subsubsection{Projects must remain modular and adaptable}
Instructors needed flexibility to adjust the scope, difficulty, and emphasis for different courses and student populations. In response, projects were revised to support alternative implementation paths, optional extensions, and clearer guidance on how to tailor the work for different contexts.

\subsubsection{Support materials are essential}
Both instructor support and student support proved essential to project success. This led to stronger documentation, setup guidance, examples, troubleshooting resources, instructor notes, and recommendations for TA and instructional support.

\subsubsection{Teamwork and formative support need structure}
Project success depended not only on project design but also on how collaboration and feedback were organized. This led to clearer guidance on team roles, accountability, checkpoints, office hours, and other ways that support timely feedback.

\subsubsection{Technical challenges must be reduced}
Outdated tools, installation problems, and platform conflicts could overshadow learning. In response, revisions emphasized more stable development environments, updated tools where possible, and earlier support for installation and configuration.

\subsubsection{AI use should be addressed in project design}
Feedback showed that students increasingly use tools such as ChatGPT during project work. Rather than prohibiting this practice, projects should be designed so that students must evaluate, critique, and improve AI-generated outputs. Projects were revised to emphasize justification of decisions, comparison of alternative solutions, debugging, and assessments that reward reasoning and verification.

\section{Discussion and Implications}
\label{section:discussion_implications}

Our experiences over four years of deployment across multiple universities suggest that structured packaging can meaningfully lower the barrier to PjBL adoption, but also reveal where packaging alone is not sufficient. In this work, packaging projects around learning objectives, coordinated student, instructor, and assessment support materials made adoption and adaptation more feasible across varied course settings. At the same time, successful implementation still depended on factors such as instructor familiarity with the project domain, available instructional support, and the complexity of the technical environment. These findings suggest that broader and more sustainable use of computing PjBL requires attention not only to project design, but also to the instructional and institutional conditions that shape adoption in practice.

\subsection{Implications for Instructors}
FORAP is intended to support instructors who want to design, adopt, or adapt computing PjBL projects. For adoption, learning objectives and project attributes help instructors judge whether a project fits their course, while milestones, checkpoints, and instructor support materials help them prepare for implementation and assessment. For adaptation, FORAP makes project structure explicit, modular, flexible, and separates materials intended for instructors from those intended for students, making it easier to adjust scope, modify deliverables, substitute tools, or tailor support to course goals and constraints while preserving alignment with the core learning objectives. At the same time, the deployments suggest that adoption is not simply plug-and-play. Even with structured packaging, instructors still need to make local decisions about scope, pacing, tooling, and support based on student preparation, course constraints, and available instructional capacity.

\subsection{Implications of the Multi-Scale Portfolio}
The multi-scale portfolio shows that a shared packaging structure can be applied across different computing domains, project scopes, and both greenfield and brownfield projects. This increases the likelihood that instructors can find a project that fits their course context, while the shared FORAP structure makes projects easier to compare, adopt, and adapt. The portfolio also serves as a set of examples for instructors who want to design additional projects using the same FORAP packaging. At the same time, the portfolio suggests that project types are not equally easy to adopt. More self-contained and shorter projects can provide lower-barrier entry points for instructors, whereas brownfield and more technically complex projects may offer a richer real-world context but typically require additional scaffolding, support, and careful scoping.

\subsection{Challenges and Future Directions}
Several challenges remain. Instructional support availability continues to affect feasibility, especially in larger classes and in technically complex projects. Even well-packaged projects require instructors and TAs to have a strong understanding of the project and its underlying technologies to provide effective support and sustain student progress. Installation and configuration challenges, along with variation in tools and platforms, can still consume substantial time, which reinforces the need for stable environments and clear installation guidance. Brownfield projects can offer realistic and valuable experiences, but they often require additional scaffolding and careful scoping to match student preparation. In addition, while FORAP includes instructor notes for sustaining a project, materials still require ongoing maintenance as dependencies, software versions, platforms, and technologies change. This is especially important for installation and configuration materials, which can become outdated quickly.

A further challenge is understanding how different support materials affect student outcomes. Prior work has explored methods for studying the effects of technical support on student success~\cite{suleiman2024exploratory}, but better approaches are still needed to isolate the effects of support materials across varied instructors, course contexts, project types, and student populations. Future work should also include broader surveys and interviews with instructors to better understand how FORAP supports adoption and adaptation across computing courses.

An emerging challenge and opportunity is the increasing availability of AI-based tools that can assist with many of the technical tasks involved in these projects, such as coding, debugging, and system integration. While these tools can reduce some barriers to project completion, they also raise questions about how to design projects and assessments that continue to promote deep understanding and skill development. Future work should explore how FORAP-based projects can be adapted to support meaningful learning in contexts where students have access to AI-assisted development tools. AI tools may also support instructors during project preparation and maintenance by helping draft or revise support materials. However, these materials still require instructor review for technical correctness and alignment with learning objectives.

A limitation of this paper is that it does not report a controlled quantitative evaluation of student outcomes or instructor workload. This paper instead focuses on presenting FORAP, the project portfolio, and implementation lessons from classroom trials. Future work will analyze student outcome data collected through the pre/post surveys and examine instructor-facing measures such as project preparation time.

\section{Conclusion}
\label{section:conclusion}

This innovative practice paper presents a multi-scale portfolio of 14 adoption-ready computing PjBL project packages and FORAP, the framework used to organize and package them for adoption and adaptation. Experiences from 44 classroom trials in seven universities over four project years provide evidence that the portfolio of adoption-ready packaging can support project adoption, implementation, and adaptation while helping to address common instructor and student challenges. 

As computing programs seek scalable, reusable, adaptable, and sustainable ways to integrate computing PjBL projects, structured packaging frameworks like FORAP offer a practical path forward by organizing projects around coordinated support for instructor, student, and assessment.

\section*{Acknowledgment}
This work was partially supported by the U.S. National Science Foundation Awards DUE-2111318, DUE-2515174, DGE-2513110, and DGE-2302615.

\balance

\bibliographystyle{IEEEtran}
\bibliography{main}

\end{document}